\documentclass[
reprint,
amsmath,amssymb,
aps,prl
]{revtex4-1}

\usepackage{graphicx}
\usepackage{dcolumn}
\usepackage{subfigure}
\usepackage{hyperref}
\usepackage{siunitx}
\usepackage{changes}

\usepackage{amssymb}
\usepackage{bm}
\usepackage{url}
\usepackage{color}

\begin{document}

\title{Unraveling the secrets of turbulence in a fluid puff}

\author{Andrea Mazzino}
\email{andrea.mazzino@unige.it}
\affiliation{Department of Civil, Chemical and Environmental Engineering (DICCA), University of Genova, Via Montallegro 1, 16145 Genova, Italy; INFN, Genova Section, Via Montallegro 1, 16145 Genova, Italy}
\author{Marco Edoardo Rosti}
\email{marco.rosti@oist.jp}
\affiliation{Complex Fluids and Flows Unit, Okinawa Institute of Science and Technology Graduate University, 1919-1 Tancha, Onna-son, Okinawa 904-0495, Japan}

\begin{abstract}
Turbulent puffs are ubiquitous in everyday life phenomena. Understanding their dynamics is important in a variety of situations ranging from industrial processes to pure and applied science. 
In all these fields, a deep knowledge of the statistical structure of temperature and velocity space/time fluctuations is of paramount importance to construct models of chemical reaction (in chemistry),  of condensation of virus-containing droplets (in virology and/or biophysics), and optimal mixing strategies  in industrial applications. As a matter of fact, results of turbulence in a puff are confined to bulk properties (i.e. average puff velocity and typical decay/growth time) and dates back to the second half of the 20th century.  There is thus a huge gap to fill to pass from bulk properties to two-point  statistical observables. Here we fill this gap exploiting theory and numerics in concert to predict and validate the space/time scaling behaviors of both velocity and {temperature} structure functions including intermittency corrections. Excellent agreement between theory and simulations is found. {Our results are expected to have profound impact to develop evaporation models for virus-containing droplets carried by a turbulent puff, with benefits to the comprehension of the  airborne  route of virus contagion.}
\end{abstract}

\maketitle 

Turbulent puffs occur whenever a fluid is impulsively ejected from a localized source in an undisturbed environment. Once the source is switched off, the cloud freely evolves into the ambient. In the freely evolving regime, the fluid cloud is named puff, to distinguish it from the initial jet phase \cite{Gh10}. Puff turbulence is a relevant example of non-ideal turbulence,  i.e.\ all space/time symmetries of ideal turbulence do not hold. It is thus a challenging playground for theoreticians to develop a theory through which characterize the statistical structure of turbulence. It is however a challenging problem also on the side of numerical analysis: the high Reynolds number characterizing the majority of interesting expulsion phenomena makes the problem technically difficult.  In addition to the interest from a fundamental perspective,  a deep, quantitative, understanding of puff turbulence nowadays emerges as a key issue of practical importance. Indeed, there are now robust evidences that the COVID-19 pandemic is largely caused by airborne transmission of virus-containing saliva droplets transported by the  puffs of human exhalations \cite{Le20,Mi20,Se20}. {Puffs are also common in industrial processes (a puff is the product of a variety of atomizers used, e.g., for disinfection/sanitizing purposes), in environmental sciences (a puff is a cloud of pollutants emitted, e.g., from chimneys), and in chemistry (a puff is a laboratory for dangerous chemical reactions, e.g., in the smoke of cigarettes) and many others.}

Despite their ubiquity, current knowledge of puff turbulence remains confined to the pioneering work by Kovasznay et al. (1975) \cite{Ko75} having as a main focus only bulk quantities involved in the puff dynamics, such as the puff bulk translational velocity and the average puff radius. Understanding the small-scale structure of fluctuations (clearly detectable in Fig.~1) still remains elusive.  Here we fill the existing gap in the present state of knowledge by proposing a statistical theory explaining the small-scale structure of turbulence fluctuations in a puff. Our theory is based on an adiabatic generalization of the Kolmogorov-Obukhov picture of steady Navier-Stokes turbulence \cite{Ko41,Ob41} where intermittency corrections are also included. The theory gives meaning to the concept of inertial range, energy flux and scaling behaviors of proper statistical observables in both space and time. The resulting predictions are compared against state-of-the-art numerical simulations of the fully-resolved puff evolution showing excellent agreement. {Predictions based on adiabatic generalizations of steady Navier-Stokes turbulence have been successful also in describing inertial range scaling laws in decaying homogeneous anisotropic turbulence \cite{Bi03}.}
\begin{figure}[t]
\centering
\includegraphics[width=0.45\textwidth]{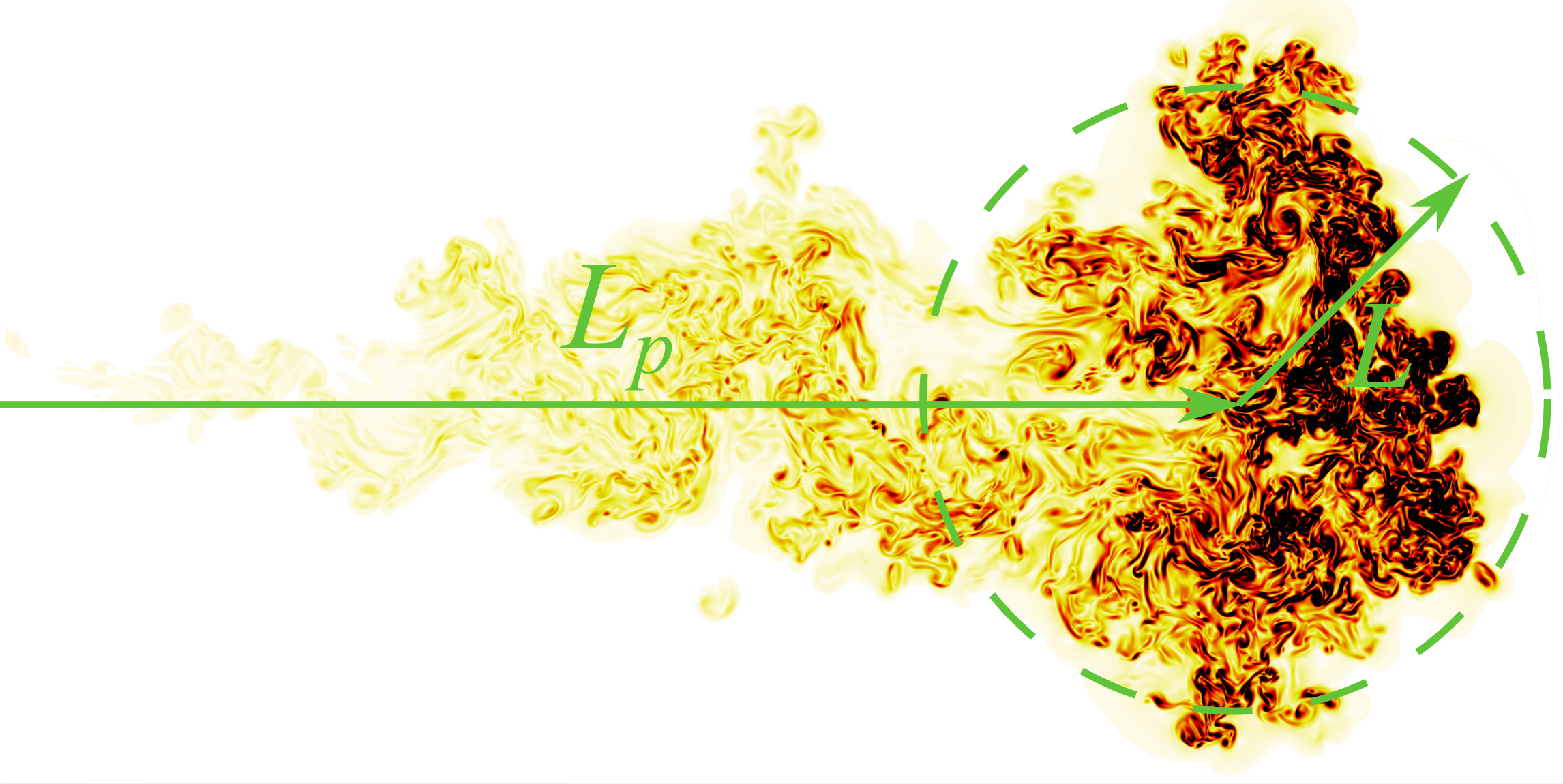}
\caption{Side view of a puff at $t/t_0 \approx 5$ {(corresponding to $Re = L \left( t \right) u_L \left( t \right) /\nu \approx 4500$)} from our simulation results. The color contour shows the magnitude of the vorticity field. }
\label{fig:sketch}
\end{figure}

The dynamics of a fluid impulsively ejected in an undisturbed environment is described in terms of the well known Oberbeck-Boussinesq (OB) equations for the velocity $\bm{u}(\bm{x},t)$ and the fluid temperature excursion, $T(\bm{x},t)$, measured with respect to the (constant) ambient temperature, $\theta_a$:
\cite{Tr88}: 
\begin{equation}
  \partial_t \bm{u} + \bm{u}\cdot \bm{\partial} \bm{u}=-\frac{\bm{\partial}p}{\rho_a} + \nu \partial^2 \bm{u} -\beta \bm{g} T,   
  \label{eq:NS}  
\end{equation}
\begin{equation}
  \bm{\partial}\cdot \bm{u}=0,
  \label{eq:div0}  
\end{equation}
\begin{equation}
  \partial_t T + \bm{u}\cdot \bm{\partial} T = \kappa \partial^2 T.
  \label{eq:T}
\end{equation}
where $\rho_a$ is the (constant) ambient density, $\beta$ is the thermal expansion coefficient, and {$\bm{g}=(0,0,-g)$ the gravitational acceleration, acting perpendicularly to the flow direction.} In Eq.~(\ref{eq:T}), $\kappa$ is the thermal diffusion coefficient and $\nu\sim \kappa$ in Eq.~(\ref{eq:NS}) the kinematic fluid viscosity {\cite{PR}}.  Our focus will be on the cloud evolution after a time $t$ larger than $t_0$, {the time of the end of the injection process when the inlet velocity is set to $0$ and the cloud is let to evolve freely, i.e.\ the puff regime \cite{NO}}. 

\begin{figure}[ht]
\centering
\includegraphics[width=0.45\textwidth]{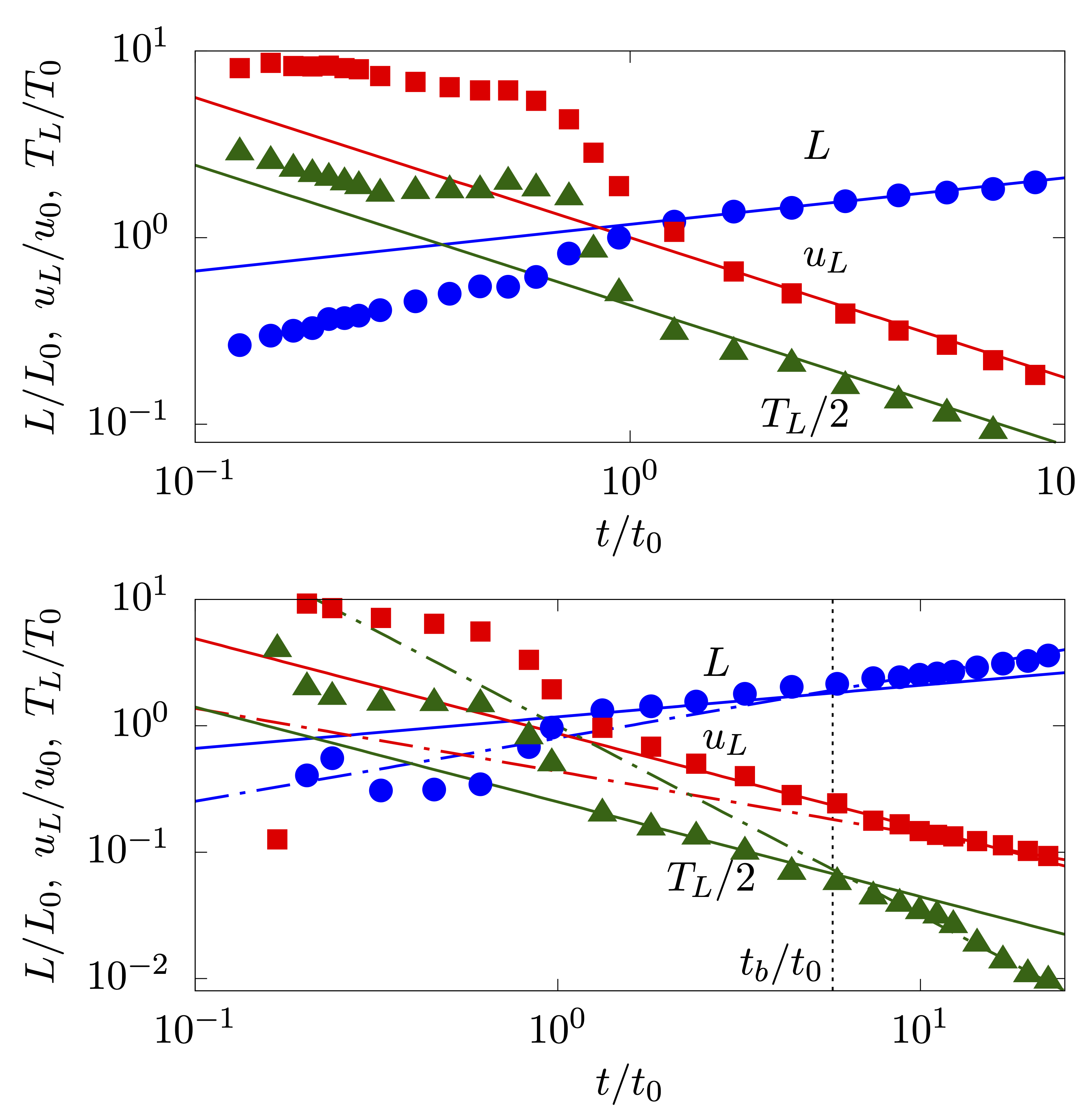}
\caption{{Scaling laws for the bulk properties of the puff $L$ (blue circle), $u_L$ (red square) and $T_L$ (green triangle, divided by a factor two for graphical reasons), for (top) the shear-induced and (bottom) the buoyancy-induced fluctuation cases with $\beta g T_0 L_0/u_0^2 \approx 0.05$ and $0.15$, respectively.  In the two panels, the solid and dash-dotted lines represent the expected scaling laws for the shear-induced and buoyancy-induced fluctuation regimes.}}
\label{fig:int}
\end{figure}

The time evolution of the bulk properties of a puff (its mean radius $L$,  its bulk velocity $u_L$, and its mean temperature $T_L$, see Fig.~1 for a sketch) can be obtained in terms of simple reasoning based on momentum conservation\cite{SM}. Their expressions read $L \sim L_0 (t/t_0)^{1/4}$, $u_L \sim  u_0 (t/t_0)^{-3/4}$ and $T_L \sim T_0 (t/t_0)^{-3/4}$, which are well-known (except the decaying law for temperature) from the seminal paper \cite{Ko75}. These scaling behaviors serve here as a benchmark for our fully-resolved numerical simulations of the puff dynamics; Fig.~2(top) shows the excellent agreement for $t/t_0 \gtrsim 1$ between the results from our simulations and the theoretical predictions.

Surprisingly, no substantial advances has been registered in the last 45 years and these results still remain the only reference predictions for puffs ejected in an initially quiescent homogeneous flow. In this Letter, we aim at filling the gap  acting on two different directions: \textit{i)} we complete the analysis of bulk properties by identifying a new scaling regime corresponding to the limit of strong buoyancy. \textit{ii)} Bulk properties constitute only a small sub-set of dynamical properties in turbulence. Here, we broaden the horizon by proposing a theory in the spirit of Kolmogorov and Obukhov for inertial-range small-scale fluctuations of both velocity and temperature.

Starting from \textit{i)}, a simple power-counting suggests that for $t\gg t_b\equiv u_0 /(\beta g T_0)$ buoyancy in Eq.~(\ref{eq:NS}) overcomes eddy-viscosity \cite{Ko75}, originating new scaling laws for $L$, $u_L$ and $T_L$ that are dictated by thermal instabilities rather than by mechanical instabilities \cite{SM}. Their expressions read $L\sim L_0  (t_0/t_b)^{1/4} (t/t_0)^{1/2}$, $u_L  \sim u_0 (t_0/t_b)^{1/4} (t/t_0)^{-1/2}$, $T_L\sim  T_0 (t_b/t_0)^{3/4} (t/t_0)^{-3/2}$. If virus-containing liquid droplets were carried by the puff as in human exhalations \cite{Ro20,Ro21,Lo21}, their condensation process would also change with impact on the contagion spread. Fig.~2-bottom shows the emergence of the new scaling law for $t \gtrsim t_b $ detected from our numerical simulations for a case having a  larger buoyancy than that reported in Fig.~2-top.

Having focused our attention on bulk properties, we now need to investigate issue \textit{ii)} on whether turbulent small-scale activity can be described in terms of classical theory \`a la Kolmogorov \cite{Ko41,Ob41}. The idea is that, despite the fact that the problem at hand is not stationary, and thus very far from the classical playground of the Kolmogorov theory, small scales of turbulence might rapidly relax to the slower large-scale dynamics. This is the essence of the adiabaticity hypothesis leading to a generalization of the classical Kolmogorov theory. Accordingly, the two-point velocity fluctuations behave as a power law in the inertial range of scales $\eta(t) \ll r \ll L(t)$, with $\eta$ being the Kolmogorov scale equal to $\eta (t)\sim \eta_0   (t/t_0)^{5/8}$ and $\eta (t)\sim \eta_0 (t_0/t_b)^{-1/8} (t/t_0)^{1/2}$ for shear-induced and for buoyancy-induced fluctuations (SIF and BIF, respectively) \cite{SM} where $\eta_0$ denotes the Kolmogorov scale at $t=t_0$, namely $\eta_0=\nu^{3/4}\epsilon_0^{-1/4}$, with $\epsilon_0$ denoting the energy flux at $t=t_0$, namely $\epsilon_0=u_0^3/L_0$.  Thus, we obtain for the two cases:
\begin{equation}
  \delta_r u (t)\sim u_0\left (\frac{r}{L_0}\right )^{1/3} \left ( \frac{t}{t_0}\right )^{-5/6} \quad \mbox{SIF},
  \label{u-nobuoy2}
\end{equation}
\begin{equation}
  \delta_r  u (t)\sim   u_0\left (\frac{r}{L_0}\right )^{1/3} \left ( \frac{t_0}{t_b}\right )^{1/6} \left ( \frac{t}{t_0}\right )^{-2/3} \quad \mbox{BIF},
  \label{u-buoy2}
\end{equation}
{where $\delta_r u $ is the velocity difference between points at distance $r$.} To arrive at Eqs.~(\ref{u-nobuoy2})-(\ref{u-buoy2}) we have supposed that buoyancy is only important to dictate the large-scale balance, being negligible within the inertial range of scales. A similar scenario has been found to hold in other convective systems as, e.g.,\ Rayleigh-Taylor turbulence \cite{Ch03,Vl09,Bo09,Ce06,Bo17}.

The viscous scaling laws (i.e.\ valid for $r \lesssim \eta$) follow by the smooth character of viscous fluctuations. Namely, from $\delta_r u \sim 
(r/\eta)\delta_{\eta} u$ one gets:
\begin{equation}
  \delta_r u (t)\sim r \left (\frac{\epsilon_0}{\nu}\right )^{1/2} \left ( \frac{t}{t_0}\right )^{-5/4} \quad \mbox{SIF},
  \label{u-viscous-nobuoy} 
\end{equation}
\begin{equation}
  \delta_r u (t)\sim r \left (\frac{\epsilon_0}{\nu}\right )^{1/2} \left ( \frac{t_0}{t_b}\right )^{1/4} \left ( \frac{t}{t_0}\right )^{-1} \quad \mbox{BIF}.
  \label{u-viscous-buoy}
\end{equation}
Because the ratio $(L/\eta)^3$ is a measure of the active degrees of freedom in a turbulent system \cite{Fr95}, from our estimations different scenarios emerge depending on whether buoyancy is important or not. In particular, while this number does not depend on time when buoyancy is dominant, it does if fluctuations are mechanically driven.

{For temperature differences between points at distance $r$, $\delta_r T $,} the adiabatic generalization of Obukhov-Corrsin theory (OC51) of passive scalar advection \cite{Ob49,Co51} yields the following inertial-range scaling laws\cite{SM}
\begin{equation}
  \delta_r T (t)\sim T_0  \left ( \frac{r}{L_0}\right )^{1/3}\left ( \frac{t}{t_0}\right )^{-5/6}\quad\mbox{SIF},
  \label{T-nobuoy2}
\end{equation}
\begin{equation}
  \delta_r T (t)\sim  T_0  \left ( \frac{r}{L_0}\right )^{1/3} \left ( \frac{t_b}{t_0}\right )^{5/6} \left ( \frac{t}{t_0}\right )^{-5/3}  \quad\mbox{BIF},
  \label{T-buoy2}
\end{equation}
accompanied by the smooth diffusive scaling behaviors, $\delta_r T \sim (r/\eta)\delta_{\eta} T$, for $r\lesssim \eta $:
\begin{equation}
  \delta_r T (t)\sim r \left ( \frac{\varepsilon_0}{\nu}\right )^{1/2} \left ( \frac{t}{t_0}\right )^{-5/4} \qquad\mbox{SIF},
  \label{T-diff-nobuoy} 
\end{equation}
\begin{equation}
  \delta_r T (t) \sim r \left ( \frac{\varepsilon_0}{\nu}\right )^{1/2} \left ( \frac{t_0}{t_b}\right )^{-3/4} \left ( \frac{t}{t_0}\right )^{-2} \qquad\mbox{BIF}.
  \label{T-diff-buoy}
\end{equation}
In all cases, we have defined $\varepsilon_0=u_0 T_0^2/L_0$ as the flux of temperature variance at $t=t_0 $.

In order to verify our theory, we need to translate the mean-field predictions for both velocity and temperature fluctuations in terms of measurable observables. The most natural statistical quantities to assess are the structure functions, i.e.\ the moments of scale-dependent fluctuations: for the velocity $S^{\parallel}_p(r)=\langle (\delta_r {\bf u}\cdot \hat {\bf r})^p  \rangle$ (longitudinal), and for the temperature $S_p(r)=\langle (\delta_r T)^p  \rangle$. Here, $\hat {\bf r}$ denotes the unit vector connecting two points separated by a  distance $r$, and brackets the average in space and over different experiment realizations. Space averages are exploited here to define isotropized/homogenized observables, from which to assess scaling behaviors as a function of the separation $r$ and time.

The simplest predictions for the scaling behaviors of the structure functions can be formulated from Eqs.~(\ref{u-nobuoy2}), (\ref{u-buoy2}) and (\ref{T-nobuoy2}), (\ref{T-buoy2}) via simple power counting. However, such a mean-field approach does not account for intermittency corrections which are typical for both velocity \cite{Fr95} and temperature fluctuations \cite{Ve97,Fr98,Sh00,Fa01,Ce01} in turbulence. Our claim here, coherently with the adiabaticity hypothesis, is that the present system possesses the same spatial scaling exponents as those of the stationary, homogeneous and isotropic turbulent system hosting a passively behaving scalar. Our predictions for a structure function of order $p$ are thus built, dimensionally, from Eqs.~(\ref{u-nobuoy2})--(\ref{u-buoy2}) and from Eqs.~(\ref{T-nobuoy2})--(\ref{T-buoy2}) with a multiplicative correction due to intermittency having the form $[r/L(t)]^{-\sigma_{p}}$ (for the velocity) and $[r/L(t)]^{-\xi_{p}}$ (for the temperature). It is indeed known that intermittency appears via the dimensionless factor $r/L$, both for the velocity fluctuations \cite{Fr95} and for the scalar fluctuations \cite{Fa01}, which cannot be captured by dimensional considerations. Interestingly, because $L$ depends on time in our case, the intermittency correction affects both the spatial and the temporal scaling laws. Because no first-principle theories are available for intermittency corrections in the stationary, homogeneous and isotropic setting, we will consider values of $\sigma_{p}$ and $\xi_p$ taken from state-of-the art numerical simulations of ideal turbulence \cite{Wa07}. As a result \cite{SM},  our model for the inertial-range behavior of the $p$-th-order structure functions becomes:
\begin{equation}
\begin{split}
  S^{\parallel}_p(r)&=A\, u_0^p \\
  &\left ( \frac{r}{L_0}\right )^{p/3-\sigma_p} \left ( \frac{t}{t_0}\right )^{-5 p/6+\sigma_p/4} \quad  \mbox{SIF},
  \label{sfu:shear.final} 
\end{split}
\end{equation}
\begin{equation}
\begin{split}
  S^{\parallel}_p(r)&=B\, u_0^p \\
  &\left ( \frac{r}{L_0}\right )^{p/3-\sigma_p} \left ( \frac{t_0}{t_b}\right )^{p/6+\sigma_p/4} \left ( \frac{t}{t_0}\right )^{-2 p /3+\sigma_p /2} \quad \mbox{BIF},
  \label{sfu:buoy.final}
\end{split}
\end{equation}
and
\begin{equation}
\begin{split}
  S_p(r)&=C \, T_0^p \\
  &\left ( \frac{r}{L_0}\right )^{p/3-\xi_p} \left ( \frac{t}{t_0}\right )^{-5 p/6+\xi_p/4} \qquad  \mbox{SIF},
  \label{sfT:shear.final}
  \end{split}
\end{equation}
\begin{equation}
\begin{split}
  S_p(r)&= D\,  T_0^p \\
  &\left ( \frac{r}{L_0}\right )^{p/3-\xi_p} \left ( \frac{t_b}{t_0}\right )^{5 p/6-\xi_p/4} \left ( \frac{t}{t_0}\right )^{-5 p/3+\xi_p/2}  \quad 
\mbox{BIF},
  \label{sfT:buoy.final}
\end{split}
\end{equation}
where A,B,C, and D are unknown non-universal constants. The corresponding structure functions in the viscous/diffusive range simply follow from Eqs.\ (\ref{u-viscous-nobuoy}) and (\ref{u-viscous-buoy}) and from Eqs.\ (\ref{T-diff-nobuoy}) and (\ref{T-diff-buoy}) without any intermittency correction.
\begin{figure}[ht]
\centering
\includegraphics[width=0.45\textwidth]{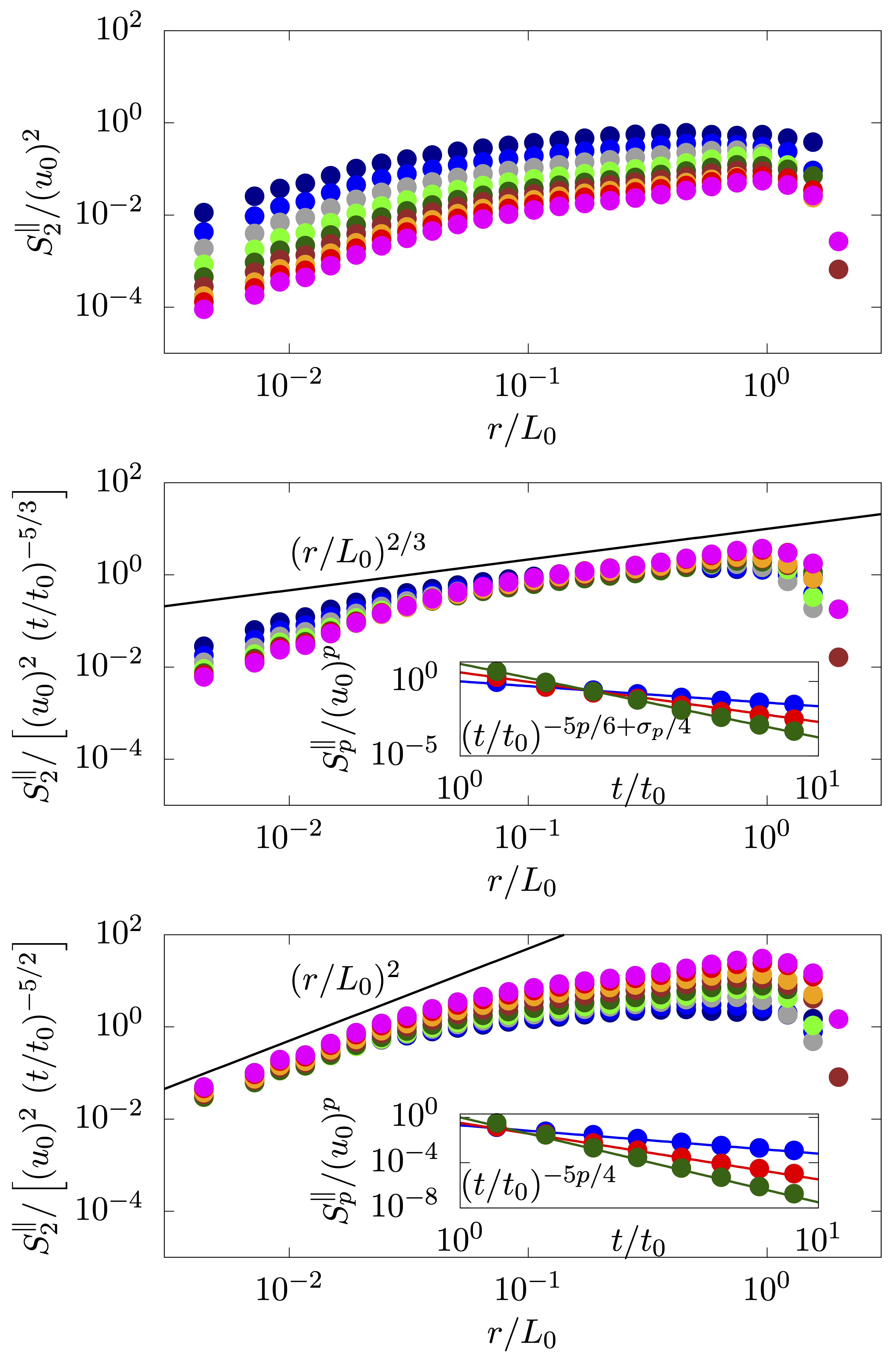}
\caption{Top: $S^{\parallel}_2(r)$ for different time instants. Middle: Same as top panel but with the ordinates scaled by the inertial-range temporal scaling law. Bottom: same as top panel but with the ordinates scaled with the viscous temporal scaling.  The insets in the middle and bottom panels report the time histories of $S^{\parallel}_2(r)$ (blue), $S^{\parallel}_4(r)$ (red) and $S^{\parallel}_6(r)$ (green) {for two separations taken in the inertial $r/L_0\approx 0.18$ (middle) and viscous $r/L_0\approx 0.007$ (bottom) range of scales.  In all the figures,  the solid lines represent the predicted slopes.  Data for the case with $\beta g T_0 L_0/u_0^2 \approx 0.05$; analogous figures for the temperature difference are reported in the Supplementary Materials \cite{SM}.}}
\label{fig:scalV}
\end{figure}
\begin{figure}[ht]
\centering
\includegraphics[width=0.45\textwidth]{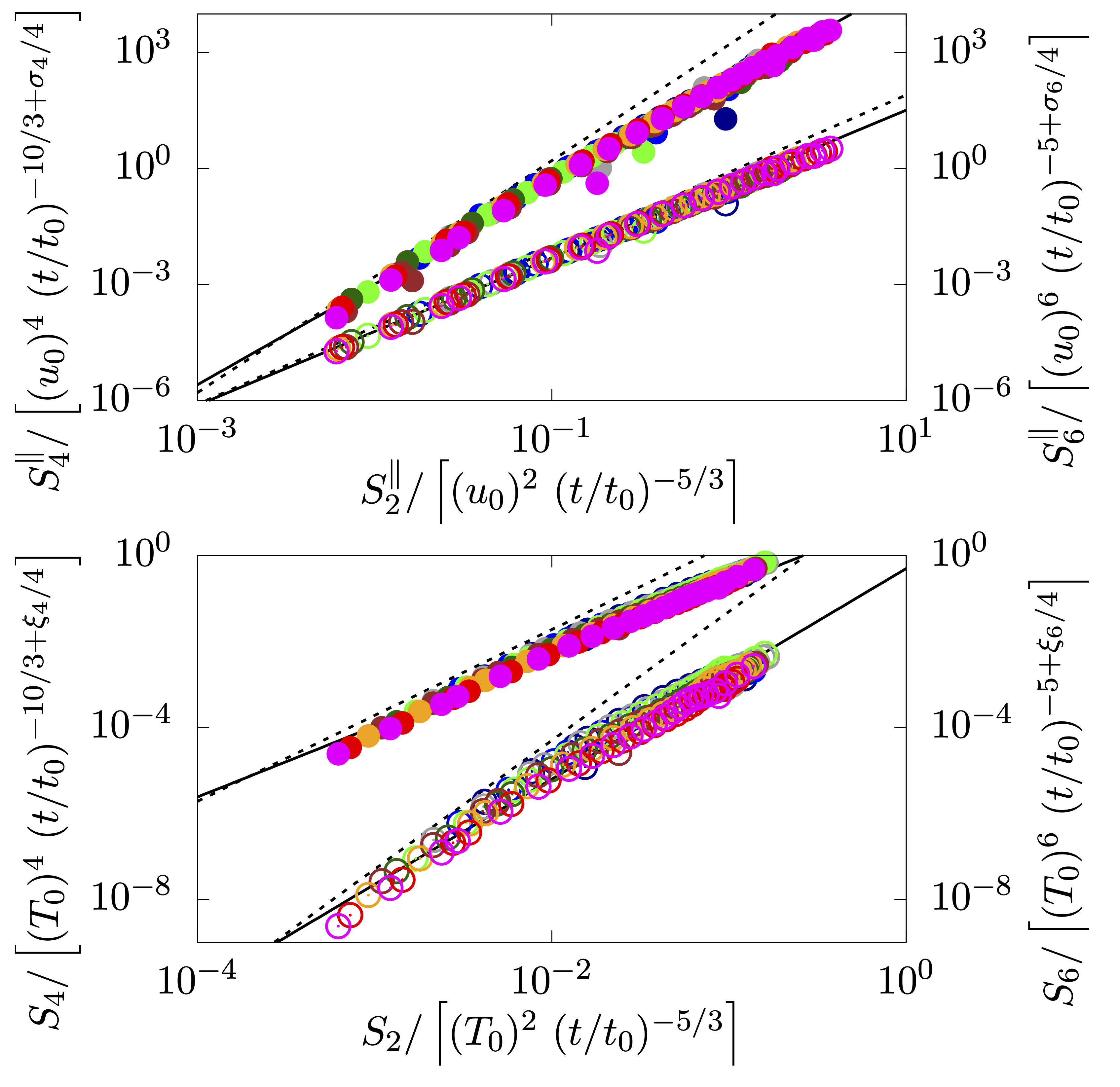}
\caption{Top: $S^{\parallel}_4(r)$ (filled symbols) and $S^{\parallel}_6(r)$ (empty symbols) as a function of $S^{\parallel}_2(r)$. The ratio of the expected scaling laws are shown with a black solid line, while the dashed lines are the same but without intermittency corrections.  In the figure, $S^{\parallel}_4(r)$ and $S^{\parallel}_6(r)$ are shifted by one decade each for the sake of graphical reasons. Bottom: same as top panel but for the temperature structure functions.  {Data for the case with $\beta g T_0 L_0/u_0^2 \approx 0.05$; analogous figures for the buoyancy dominated regime are reported in the Supplementary Materials \cite{SM}.}}
\label{fig:scal}
\end{figure}

We now verify in Figs.~3 and 4 the validity of our scaling predictions for structure functions (both in the  inertial and viscous range of scales) in terms of our high-resolution fully-resolved direct numerical simulations. Fig.~3-top shows the velocity second-order structure function,  $S^{\parallel}_2(r)$, at different time instants always larger than $t_0$. The validity of the proposed scaling laws can be verified by the excellent collapse of all the structure functions when these laws are used to normalize the data (Figs.~3-middle and 3-bottom). In particular, Fig.~3-middle shows the same data of 3-top scaled with the temporal scaling law (\ref{sfu:shear.final}) expected to hold in the inertial range of scales. The region where a clean data collapse does appear defines the universal inertial range of scales. Consistently, we  observe data spreading in the viscous range. This fact has a simple interpretation being due to the different decay laws characterizing the viscous range and the inertial range of scales. To show that this is indeed the case, in Fig.~3-bottom we have rescaled the ordinate of Fig.~3-top with the expected temporal viscous scaling (\ref{u-viscous-nobuoy}) finding data collapse in the viscous range (and not, as expected, in the inertial range).  Moreover, in the inset of Fig.~3-middle we have focused on a separation, $r$, belonging to the inertial range and analyzed the temporal behavior of the velocity structure functions of order 2, 4, and 6. The continuous lines are the corresponding scaling laws from our theoretical predictions given by (\ref{sfu:shear.final}) for orders 2, 4, and 6. The inset of Fig.~3-bottom is similar, apart from the fact that the separation belongs to the viscous range and the theoretical prediction (\ref{u-viscous-nobuoy}) is used accordingly. In both cases, an excellent agreement between theory and numerics is found, also signaling that our model correctly captures intermittency corrections. These latter indeed enter not only in the spatial scaling but also in the temporal scaling laws.

Fig.~4 completes the verification by looking at the scaling laws (spatial and temporal) of high-order velocity and temperature structure functions,  with data shown in the form of Extended Self-Similarity (ESS, see \cite{Be93}). The continuous lines have the slopes deduced from our predictions (\ref{sfu:shear.final}) and (\ref{sfT:shear.final}) while dashed lines do not account for the intermittency corrections (i.e.\ they represent a pure mean field prediction). Different times $t>t_0 $ are considered for $\beta g T_0 L_0/u_0^2 \approx 0.05$ corresponding to the shear-dominated case. Both the excellent data collapse and the convincing correspondence of our data with the theoretical slopes (continuous lines) provide a further confirmation of our predictions (\ref{sfu:shear.final})--(\ref{sfT:buoy.final}).

In conclusion, we have proposed a phenomenolgical theory for a fluid puff exhaled in an undisturbed environment. The foundation  of our theory is the adiabaticity hypothesis leading to a generalization of the classical Kolmogorov theory. Intermittency corrections have also been inferred assuming that a puff possesses the same spatial scaling exponents as the ideal Navier-Stokes turbulence system. The resulting statistical predictions for a turbulent puff have been validated against state-of-the-art numerical simulations founding an excellent agreement. {The present work can have profound implications in the accurate prediction of the airborne spreading of virus where droplet evaporation,  mainly controlled by the combined effect of turbulence and droplet inertia \cite{Ro21},  strongly affects the traveled distance of the viral load \cite{Ro20}.}

\begin{acknowledgments}
\subsection{Acknowledgments}
M.E.R. acknowledges the computational time provided by HPCI on the Oakbridge-CX cluster in the Information Technology Center, The University of Tokyo, under the grant hp200157 of the ``HPCI Urgent Call for Fighting against COVID-19'' and the computer time provided by the Scientific Computing section of Research Support Division at OIST. A.M. thanks the financial support from the Compagnia di San Paolo, project MINIERA no. I34I20000380007, and the FISR COVID-19 project FISR2020IP 00290 from the ``Ministero dell’Universit\`a e della Ricerca'' (MUR).
\end{acknowledgments}


\begin{thebibliography}{45}%
\bibitem{Gh10}  A. Ghaem-Maghami, H. Johari, Velocity field of isolated turbulent puffs. Phys Fluids 22, 115105 (2010).

\bibitem{Le20} D. Lewis, Is the coronavirus airborne? Experts can't agree. Nature 580, 175 (2020).

\bibitem{Mi20} R. Mittal, R. Ni, J. Seo, The flow physics of COVID-19. J. 
Fluid Mech. 894, F2 (2020).
  
\bibitem{Se20} G. Seminara, B. Carli, G. Forni, S. Fuzzi, A. Mazzino, A. Rinaldo, Biological fluid dynamics of airborne COVID-19 infection. Rend. Fis. Acc. Lincei 31, 505-537 (2020).

\bibitem{Ko75}  L.S. Kovasznay, H. Fujita, R.L. Lee, Unsteady turbulent puffs, in Advances in Geophysics 18, 253-263 (Elsevier, Amsterdam, 1975).

\bibitem{Bi03}  L. Biferale, G. Boffetta, A. Celani, A. Lanotte, F. Toschi, and M. Vergassola, The decay of homogeneous anisotropic turbulence. Phys Fluids 15, 2105 (2003).

\bibitem{Tr88} D. Tritton, Physical Fluid Dynamics. Oxford, UK: Clarendon 
(1988).

\bibitem{PR}  {Our choice for $Pr=\nu/\kappa=1$ follows from the fact that air thermal diffusivity and air viscosity are very close to each other. Also, when $Pr=1$ an inertial range of scales for both velocity and temperature fluctuations exist for the same energy-containing scales; this is not the case neither for $Pr>>1$, when the scalar evolves in a smooth turbulent background without intermittency, nor for $Pr<<1$, when the scalar exhibits a trivial bare diffusive regime.  As far as the puff bulk properties are concerned, $Pr$ does not play a role neither on the bulk velocity decay nor in the growing of the puff size, which are a consequence of the momentum conservation alone.}

\bibitem{NO} {In the current work,  the suffix `$_0$' is used to indicate properties evaluated at $t=t_0$.}


\bibitem{Ro20} M.E. Rosti, S. Olivieri, M. Cavaiola, A. Seminara, A. Mazzino, Fluid dynamics of COVID-19 airborne infection suggests urgent data for a scientific design of social distancing, Sci. Rep. 10, 22426 (2020).

\bibitem{SM}  See the Supplemental Materials \cite{gupta2009flow,  rosti_brandt_2017a, rosti2019flowing, rosti_ge_jain_dodd_brandt_2019,rosti2020increase,  olivieri2020dispersed}.

\bibitem{Ro21}  M.E. Rosti, M. Cavaiola, S. Olivieri, A. Seminara, A. Mazzino, Turbulence role in the fate of virus-containing droplets in violent 
expiratory events, Phys. Rev. Res. 3, 013091 (2021).

\bibitem{Lo21}  K.L. Chong, C.S. Ng, N. Hori, R. Yang, R. Verzicco, D. Lohse, Extended Lifetime of Respiratory Droplets in a Turbulent Vapor Puff and Its Implications on Airborne Disease Transmission, Phys. Rev. Lett. 126, 034502 (2021).

\bibitem{Ko41}  A.N. Kolmogorov, The local structure of turbulence in incompressible viscous fluid for very large Reynolds numbers. C.R. Acad. Sci. URSS 30, 301-305 (1941).

\bibitem{Ob41} A. Obukhov, On the distribution of energy in the spectrum of turbulent flow. C.R. Acad. Sci. URSS32, 22-24 (1941).

\bibitem{Ch03} M. Chertkov, Phenomenology of Rayleigh-Taylor turbulence. Phys. Rev. Lett. 91, 115001 (2003).

\bibitem{Bo09} G. Boffetta, A. Mazzino, S. Musacchio, L. Vozella, Kolmogorov scaling and intermittency in Rayleigh-Taylor turbulence, Phys. Rev. E 
79, 065301(R) (2009).

\bibitem{Ce06}  A. Celani, A. Mazzino, L. Vozella, Rayleigh-Taylor Turbulence in Two Dimensions, Phys. Rev. Lett. 96, 134504 (2006).
  
\bibitem{Bo17}  G. Boffetta, A. Mazzino, Incompressible Rayleigh-Taylor turbulence, Ann. Rev. Fluid Mech. 49, 119-143 (2017).

\bibitem{Vl09} N. Vladimirova, M. Chertkov, Self-similarity and universality in Rayleigh-Taylor, Boussinesq turbulence. Phys. Fluids 21, 015102 (2009).

\bibitem{Fr95} U. Frisch, Turbulence: The Legacy of A.N. Kolmogorov. Cambridge, UK: Cambridge Univ. Press (1995).

\bibitem{Ob49}  A. Obukhov, Temperature field structure in a turbulent flow. Izv. Acad. Nauk SSSR Geogr. Geophys. 13, 58-69 (1949).

\bibitem{Co51} S. Corrsin, On the spectrum of isotropic temperature fluctuations in an isotropic turbulence. J. Appl. Phys. 22, 469-473 (1951).

\bibitem{Sh00} B.I. Shraiman, E.D. Siggia, Scalar turbulence, Nature 405, 
639-646 (2000).

\bibitem{Fr98}  U. Frisch, A. Mazzino, M. Vergassola, Intermittency in passive scalar advection, Phys. Rev. Lett. 80, 5532-5537 (1998).

\bibitem{Ve97} M. Vergassola, A. Mazzino, Structures and intermittency in 
a passive scalar model, Phys. Rev. Lett. 79, 1849-1852 (1997).

\bibitem{Fa01} G. Falkovich, K. Gawkedzki, M. Vergassola, Particles and fields in fluid turbulence, Rev. Mod. Phys. 73, 913-975 (2001).

\bibitem{Ce01} A. Celani, A. Mazzino and M. Vergassola, Thermal plume turbulence, Phys. Fluids 13, 2133-2135 (2001)
  
\bibitem{Wa07} T. Watanabe,  T. Gotoh, Inertial-range intermittency and accuracy of direct numerical simulation for turbulence and passive scalar turbulence. J. Fluid Mech. 590, 117-146 (2007).

\bibitem{Be93} R. Benzi, S. Ciliberto, R. Tripiccione, C. Baudet, S. Massaioli, S. Succi, Extended self-similarity in turbulent flows. Phys. Rev. E 48, R29(R) (1993).
 
\bibitem{gupta2009flow} J.K. Gupta, C.H. Lin, Q. Chen, Flow dynamics and characterization of a cough. Indoor Air, 19(6):517–525 (2009).

\bibitem{rosti_brandt_2017a} M.E. Rosti, L. Brandt, Numerical simulation of turbulent channel flow over a viscous hyper-elastic wall. Journal of Fluid Mechanics 830, 708–735 (2017).

\bibitem{rosti2019flowing} M.E. Rosti, S. Olivieri, A.A. Banaei, L. Brandt, A. Mazzino, Flowing fibers as a proxy of turbulence statistics. Meccanica 55, 357–370 (2020).

\bibitem{rosti_ge_jain_dodd_brandt_2019} M.E. Rosti, Z. Ge, S.S. Jain, M.S. Dodd, L. Brandt, Droplets in homogeneous shear turbulence. Journal of Fluid Mechanics 876, 962–984 (2019).

\bibitem{rosti2020increase} M.E. Rosti, L. Brandt, Increase of turbulent drag by polymers in particle suspensions. Physical Review Fluids 5, 041301(R) (2020).

\bibitem{olivieri2020dispersed} S. Olivieri, L. Brandt, M.E. Rosti, A. Mazzino, Dispersed fibers change the classical energy budget of turbulence via nonlocal transfer. Physical Review Letters 125, 114501 (2020).

\end{thebibliography}

\end{document}